\documentclass[12pt,epsf]{article}
\usepackage{epsfig}
\usepackage{cite}

\begin{document}

\title{Two-photon processes in faint biphoton fields}
\author{Dmitry V. Strekalov, Matthew C. Stowe, Maria V. Chekhova$^*$ \\
and Jonathan P. Dowling.\\
Quantum Computing Technologies Group, Sect. 367,\\
Jet Propulsion Laboratory, California Institute of Technology,\\
MS 300-123, 4800 Oak Grove Drive, Pasadena, CA 91109.\\
Dmitry.V.Strekalov@jpl.nasa.gov\\
$^*$Moscow State University, \\
Moscow, Russia}
\maketitle

\begin{abstract}
The goal of this research is to determine and study a physical system that will enable a fast and intrinsically two-photon detector, which would be of interest for 
quantum information and metrology applications. We consider two types of two-photon processes that can be observed using a very faint, but quantum-correlated 
{\em biphoton} field. These are optical up-conversion and an external photoelectric effect. We estimate the correlation enhancement factor for the biphoton light 
compared to coherent light, report and discuss the preliminary experimental results.
\end{abstract}
\newpage

\section{Introduction}

The term ``biphoton" has been suggested by D.N. Klyshko \cite{klyshko82} for describing systems of entangled photon pairs, such as those produced in the 
Spontaneous Parametric Down Conversion (SPDC) of light\footnote{In acknowledgment of D. N. Klyshko's founding role in studies of parametric processes in 
optics, SPDC is sometimes referred to as ``Klyshko radiation".}. In this process, a pump photon is coherently converted into a pair of entangled photons  in a 
$\chi^{(2)}$ medium \cite{Giord65}, while the phase and spectral properties of the pump are transferred to the biphoton \cite{burlakov97,burlakov01}. This 
phase then can be recovered by a two-photon correlation measurement. In this context, describing this system as ``two photons" is not complete and even may be 
misleading as it tends to omit the quantum correlation properties of the photon pair \cite{pittman96a,strekalov98}; so the term ``biphoton" proves to be very useful.

A pair of entangled particles that are space-like separated is an excellent system for testing the concepts of physical \emph{locality} and \emph{reality} 
\cite{einstein35}. Therefore it is not surprising that SPDC has received great attention as an entanglement source for Bell inequalities tests. The whole variety of such 
tests can be generally described as nonlocal biphoton interference experiments. As a further development, experiments involving higher-order interference have been 
carried out, such as the Greenberger-Horne-Zeilinger theorem test \cite{pan00ghz} and quantum teleportation \cite{Bouwmeester97teleport,kim01teleport}.

Besides purely a fundamental interest, study of the biphoton light has inspired development of new information and measurement technologies, such as quantum 
cryptography \cite{ekert91a}, sub-shot noise optical phase measurements \cite{Yuen86,Yurke86,lugiato02qimaging}, quantum clock synchronization 
\cite{jozsa00clock,Giovannetti02clock}, and others. Most recently, the possibility of sub-diffraction-limited imaging \cite{lugiato02qimaging} and lithography 
\cite{boto00,kok01,Bjork01,dangelo01,strekalov02litho} has been discussed. There is a clear shift  of the research in biphoton optics towards practical 
applications. It is also clear that the difficulties on the path to such applications are quite severe and have one common root, which is the detection problem.

Consider quantum cryptography, which is generally believed to be the most mature of quantum optical technologies. It has been shown \cite{gilbert00} that the main 
limitation blocking its practical implementation is the single-photon detection rate. It is not possible, with the state-of-art single-photon detectors, to achieve the 
detection rate that would make useful wideband quantum crypto-communication channels \cite{gilbert00}. Similarly, quantum optical interferometers are 
theoretically capable of being noise-limited at the fundamental (Heisenberg) level $1/<n>$, while the classical shot-noise limit is $1/\sqrt{<n>}$ ($<n>$ being the 
mean number of photons per measurement). Therefore the quantum measurement of the phase has advantage over the classical one only when $<n>$ is large. But 
for a large photon flux, any coincidence detection technique will fail because of the finite dead time of photon counting detectors and coincidence circuit 
time-windows. Thus again, the detection rate of photon pairs is the limiting factor. The situation gets even more complicated for quantum lithography, where the 
two-photon detectors are molecules in the resists that are small and distributed in the media. Here one has to worry not only about the detection rate, or fast 
two-photon sensitivity, but also about the transverse correlation properties of the biphoton light.

Therefore the search for fast two-photon processes that could enable a true two-photon detector (as opposed to an electronic coincidence circuit), that could be 
used for study of strong biphoton fields, becomes one of the most important questions of the applied quantum optics. So far, no two-photon processes induced by 
the biphoton field have been demonstrated, although a much brighter radiation consisting of photon pairs from optical parametric oscillator (OPO) has been shown 
to be more efficient for second harmonic generation \cite{abram86} and for two-photon ionization of atoms \cite{georgiades95,georgiades99} than the ordinary 
coherent radiation. On the other hand, two-photon processes in strong classical fields, such as pulsed laser fields, have been studied quite extensively.

The goal of our research is to analyze the possibility of direct two-photon detection of the biphoton field and to determine the physical system which is most suitable 
for this purpose. In the following sections we will demonstrate the advantage of biphoton light over coherent light for driving two-photon processes. This advantage 
will be shown considerable enough to make up for extremely low power of the biphoton sources such as SPDC. Then, we will discuss two types of processes that 
could serve for direct two-photon detection: optical up-conversion and the photo-electric effect.

\section{Photon detection in biphoton and coherent fields}

We will use a conventional concept of a photon detection event \cite{perinabk}, which takes place with a probability proportional to a constant $\eta$ (the 
\emph{quantum efficiency} of the detector) and to the probability to have at least one photon in the detection volume $V$. This volume can be generally defined in 
phase space as $V=\Delta k_x\Delta k_y\Delta k_z\Delta x\Delta y\Delta z$, just like the phase-space volume occupied by radiation from a source. For example, in 
the $(x,p_x)$ subspace of the phase space, a point-like detector may be represented as a vertical line, whereas a distributed but wavelength selective detector (a 
bandpass filter) will be represented as a horizontal line. If the source or the detector are single-mode, then their x- and k-volumes are reciprocal to each other, and 
$V=(2\pi)^3$. Therefore the number of modes a detector can see, or a source can radiate, can be calculated as
\begin{equation}
M=\frac{V}{(2\pi)^3}=\frac{AL}{(2\pi)^3}\frac{k^2}{c}\Delta\Omega\Delta\omega,
    \label{M}
\end{equation}
where for the x-part of $V$ we substitute a product of the detector's cross section $A$ and its length $L$, and for the k-part we take
\begin{equation}
\Delta k_x\Delta k_y\Delta k_z=k^2\Delta\Omega\Delta k=\frac{k^2}{c}\Delta\Omega\Delta\omega.
    \label{Vk}
\end{equation}
In Eqs. (\ref{M}) and (\ref{Vk}), $\Delta\Omega$ and $\Delta\omega$ are the solid angle and frequency ranges, respectively, and $c$ is the speed of light.

\begin{figure}[htp]
\centerline{
\input epsf
\setlength{\epsfxsize}{5in}
\setlength{\epsfysize}{3in}
\epsffile{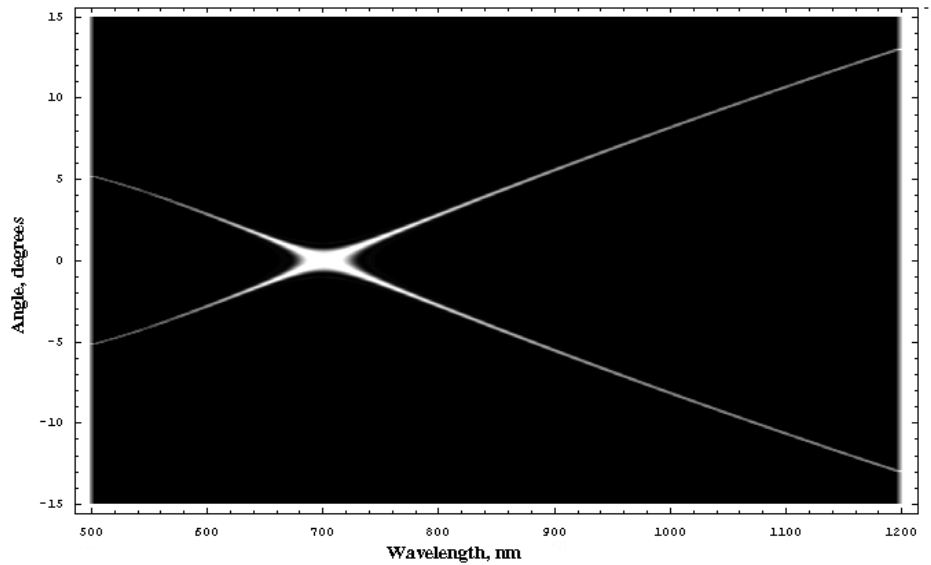}
}
\caption[setup]{\label{fig:spdcranges}The SPDC spectrum for a 5 mm Type-I BBO cut for the collinear degenerate phase matching.}
\end{figure}

Knowing the definition of the ``detection event" and associated with it phase-space volume, we can estimate detection rate for radiation with known properties. For 
example, in the next section we will consider optical up-conversion process, which is the reverse of the SPDC process, as a two-photon detector. A peculiarity of 
such a detector is that the phase space volumes for two up-converting photons are different. Any \emph{signal} photon can be up-converted anywhere within the 
illuminated part of the crystal (the x-volume) and within the k-volume whose ($\lambda,\theta$) section is shown as a tuning curve  in Fig.\ref{fig:spdcranges}. But 
the up-conversion only may take place in presence of the \emph{idler} photon satisfying the phase matching conditions with the signal photon:
\begin{eqnarray}
\omega_{s}+\omega_{i}&=& \omega_p,\nonumber \\
\vec{k}_{s}+\vec{k}_{i} &=& \vec{k}_p, \label{phasematch}
\end{eqnarray}
where $\omega_p$ and $\vec{k}_p$ are the frequency and wave vector of the pump. The choice which photon is the signal and which one is the idler is arbitrary, 
just like in SPDC.

The phase space volume for a signal photon contains a large number $M$ of modes, however the phase space volume for the \emph{phase-matching} idler photon 
is transform-limited in all directions for a single-mode pump. Hence this volume is equal to $(2\pi)^3$ and contains only one mode. In the limit of weak fields, 
$<n>\ll 1$, both volumes are filled with at least one photon each with probability $M<n>\times<n>$. When this condition is fulfilled, a detection occurs with 
probability $\eta^{(2)}$ which absorbs the quantum efficiencies of non-linear up-conversion and of subsequent photo detection. Therefore the two-photon 
detection probability from a coherently illuminated up-conversion detector can be estimated as
\begin{equation}
R_{coh} = \eta^{(2)}M<n>^2.
    \label{Sc}
\end{equation}

Even though the detector we describe is capable of seeing a large number $M_{spdc}$ of modes, not all of them may be filled by the radiation source. In practice, 
the laser sources used in the up-conversion and second harmonic generation experiments are well collimated and have a relatively narrow frequency band width. 
Therefore in Eq. (\ref{Sc}), we should take the number of modes of a laser source $M=M_{laser}$.

Now instead of coherent light let us consider the biphoton light. The most important distinction is that under proper imaging conditions (more detail in the next 
section), for every signal photon the phase-matched idler photon is guaranteed to be present, so the detection volumes are filled with at least one photon each with 
probability $M<n>$. In contrast with Eq. (\ref{Sc}), in this case $M=M_{spdc}$, and instead of (\ref{Sc}) we get
\begin{equation}
R_{spdc} =\eta^{(2)}M_{spdc}<n>.
    \label{Sspdc}
\end{equation}

Comparing the two-photon signal from coherent light (\ref{Sc}) and from biphoton light (\ref{Sspdc}) we can determine how much biphoton light is more efficient 
than coherent light for the type of two-photon process we presently consider. This comparison can be carried out either for equal intensities, or for equal $<n>$. 
We will carry it out for equal intensities, since this will allow for a direct comparison with known experiments.

The mean number of photons per mode $<n>$ can be expressed in terms of brightness as \cite{klyshkobk}
\begin{equation}
<n>=\frac{I\lambda^3}{\hbar c}\frac{1}{\Delta\Omega\Delta\omega},
    \label{n}
\end{equation}
where $I$ is intensity. Multiplying (\ref{n}) by (\ref{M}) we get the following equation:
\begin{equation}
M<n>=\frac{IAL}{c\hbar\omega},
    \label{nM}
\end{equation}
which appears rather obvious, since its left side is the mean total number of photons, and the right side is the total energy of radiation divided by the energy of a 
photon.

Now we can directly compare the signals from coherent and biphoton fields:
\begin{equation}
\xi\equiv\frac{R_{spdc}}{R_{coh}}=\frac{M_{spdc}<n>_{spdc}}{M_{coh}<n>_{coh}^2}.
    \label{ratio}
\end{equation}
Substituting (\ref{n}) and (\ref{nM}) in (\ref{ratio}) and assuming $I_{spdc}=I_{coh}\equiv I$, we find
\begin{equation}
\xi=\frac{\hbar c\Delta\Omega_{coh}\Delta\omega_{coh}}{I\lambda^3}.
    \label{ratio1}
\end{equation}

Now let us make numeric estimates. We consider a Type-I BBO crystal near collinear degenerate phase matching with the central wavelength $\lambda_0 = 702$ 
nm, whose spectrum is shown in Fig.\ref{fig:spdcranges}. The typical light power we get from a 5 mm - long Type-I BBO crystal is 50 nW. This SPDC light is 
generated by the pump beam that is approximately 100 microns in diameter. This gives us $I\approx 5 \,W/m^2$. For comparison, we will take a mode-locked 
pulsed laser with pulse width of 150 fs ($\Delta\omega_{coh}\approx 4\times 10^{13}\,s^{-1}$) and duty cycle of $10^{-5}$. This laser is focused into a 100 
$\mu m$ spot with near-diffraction divergence of $\theta_d\approx\lambda/d\approx 5\times 10^{-5}$ radians, which gives us $\Delta\Omega\approx 
2\pi\theta_d^2\approx 3\times 10^{-4}$ st. radians. Substituting these numbers into Eq. (\ref{ratio1}) we obtain $\xi\approx 200$. It is easy to see that an estimate 
carried out for $<n>_{spdc}=<n>_{coh}$ (and valid when both types of radiation have similar spectral width and divergence) gives a much larger value of $\xi$.

\section{Up-conversion of biphoton field}

\begin{figure}[htp]
\centering
\includegraphics[width=5.7in]{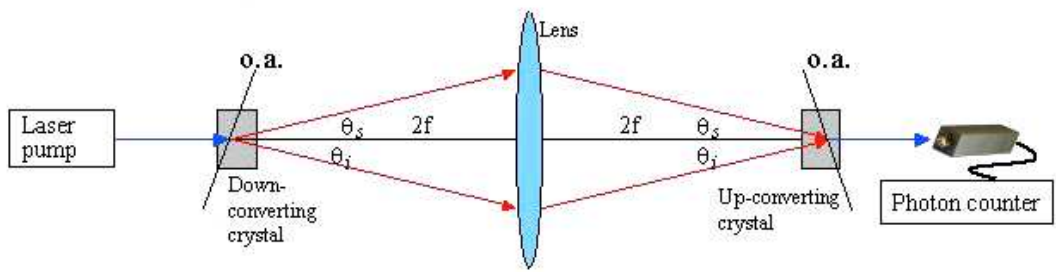}
\caption{A conceptual drawing of the experiment on up-conversion of biphoton light. O.A. stands for the optical axes of the nonlinear crystals.} 
\label{upconversion}
\end{figure}

Up-conversion of strong coherent fields is routinely used in optical autocorrelators to measure the duration of short pulses. The large value of the enhancement factor 
$\xi$ for the biphoton light suggests that the optical up-conversion can be observed with such light, despite a typically very low efficiency of this process. It should 
be possible, using a pair of similar nonlinear crystals (the first one for SPDC and the second one for the up-conversion), to detect the up-converted UV photons 
emitted from the second crystal. Fig.\ref{upconversion} is a conceptual drawing of a system which brings back together all photon pairs emitted from a localized 
source at various angles, while preserving the phase-matching conditions (that are equivalent for SPDC and for the up-conversion) by an imaging system with a unity 
magnification. This system is represented by a lens placed between the two nonlinear crystals at the distance $2f$ from each.

A rough estimate of the expected detection rate of the up-converted UV photons can be carried out based on the above-found biphoton enhancement factor and on 
the experimental observation that a $1$ Watt laser pump such as described in the previous section focused down to about 100 microns on a BBO crystal similar to 
the one we use for up-conversion produces approximately  600 mW of the second harmonic. Scaling this number quadratically to a 50 nW pump power and 
multiplying by the duty cycle to obtain an estimate for CW, we get $1.5\times 10^{-20}$ W of the up-converted light, or about 0.2 photons per second. To take 
the biphoton nature of our ``pump'' into account, we multiply this number by the enhancement factor $\xi$, arriving at $40$ photons per second. Such a signal 
should be detectable with a low-noise photomultiplier tube with a few percent quantum efficiency. However, this number gets significantly lower if a more careful 
evaluation is carried out, and various mechanisms of the phase mismatch are taken into account. To perform such an evaluation we need to understand the 
transverse correlation properties of a biphoton.

While the longitudinal (temporal) correlation in the biphoton field has been extensively studied in both theory and experiment, starting from the pioneering works 
\cite{shih88,hong87}, the studies of its transverse correlation are few. A very broad theoretical description is given in \cite{rubin96}; theoretical analysis along with 
experimental data is also presented in \cite{burlakov97}; study of biphoton propagation through generic imaging systems is reported in 
\cite{abouraddy01imaging,nasr02biphoton}.

Summarizing the analysis from \cite{burlakov97}, we write the quantum state of a biphoton emitted in a monochromatically pumped SPDC process as
 \begin{equation}
|\Psi\rangle=\int F(\vec{k}_s,\vec{k}_i)|1\rangle_{\vec{k}_s}|1\rangle_{\vec{k}_i}\, d\vec{k}_sd\vec{k}_i,
    \label{psi}
\end{equation}
where the delta-like two-photon amplitude $F(\vec{k}_s,\vec{k}_i)$ which entangles  the signal SPDC mode labelled by its wave vector $\vec{k}_s$ with the 
idler mode $\vec{k}_i$ breaks up into a product of its longitudinal and transverse parts:
 \begin{equation}
F(\vec{k}_s,\vec{k}_i)=
F_x(\Delta\omega,\Delta\theta_s,\Delta\theta_i)F_z(\Delta\omega
,\Delta\theta_s,\Delta\theta_i).
    \label{f}
\end{equation}
In Eq. (\ref{f}), the arguments of the two-photon amplitude $F$, representing the wave vectors and frequencies of the signal and idler photons, are first replaced by 
the variations of these values from their central values that satisfy the phase matching conditions (\ref{phasematch}) so that
\begin{eqnarray}
\Delta\omega &\equiv& \omega_{s}({\vec{k}_{s}})-\omega_{s0}({\vec{k}_{s0}})=\omega_{i0}({\vec{k}_{i0}})-
\omega_{i}({\vec{k}_{i}}),\nonumber \\
\Delta\vec{k}_s &\equiv&\vec{k}_{s}- \vec{k}_{s0},\\
\Delta\vec{k}_i &\equiv&\vec{k}_{i}- \vec{k}_{i0}.\nonumber
    \label{deltas}
\end{eqnarray}
Then, the longitudinal ($z$) and transverse ($x$) components of the phase mismatch $\Delta\vec{k}_{s}+\Delta\vec{k}_{s} $ are expressed via $\Delta\omega$ 
and the variations of the angles $\Delta\theta_{s,i}$ to result into the expressions for the $F_x$ and $F_z$ in Eq. (\ref{f}).

The physical meaning and the use of the two-photon amplitude (\ref{f}) is quite transparent. One can set $\Delta\theta_{s,i}=0$ and study only the temporal, or 
longitudinal, correlation of the biphoton. This is often done since such an approximation describes the most usual type of experiments with SPDC, when the signal 
and idler modes are selected by a set of narrow pinholes. On the other hand, one could collect the biphoton light from a broad range of angles, but through a pair of 
very narrow band-pass optical filters. This case corresponds to the limit $\Delta\omega=0$ which gives the angular, or transverse, part of the biphoton amplitude, 
which we need to find.

As it has been shown in \cite{burlakov97}, $F_x$ reflects the properties of the pump angular spectrum and the transverse inhomogeneities of the crystal; it has a 
Gaussian shape for a Gaussian shaped pump. $F_z$ depends on the polarization and wavelength dispersion properties of the nonlinear crystal. The analysis for a 
monochromatic Gaussian pump of the diameter $a$ and for a Type-I crystal of the length $l$ yields simple expressions for $F_x$ and $F_z$ as functions of the 
internal angles $\theta^{(in)}$. Assuming $\Delta\omega=0$ we have
\begin{eqnarray}
F_x &=& \exp\left\{-\frac{(2\pi a)^2}{4}\left(\frac{n(\lambda_s)}{\lambda_s}\cos[\theta^{(in)}(\lambda_s)]\Delta\theta
^{(in)}_s - \frac{ n(\lambda_i)}{\lambda_i}\cos[\theta^{(in)}(\lambda_i)]\Delta\theta^{(in)}_i\right)^2
\right\},\nonumber \\
F_z &=& {\rm sinc}\left\{-\frac{2\pi l}{2}\left(\frac{n(\lambda_s)}{\lambda_s}\sin[\theta^{(in)}(\lambda_s)]
\Delta\theta^{(in)}_s + \frac{ n(\lambda_i)}{\lambda_i}\sin[\theta^{(in)}(\lambda_i)]\Delta\theta^{(in)}_i\right)
\right\}. \label{fxfz}
\end{eqnarray}
The wavelengths $\lambda_{s,i}$ correspond to ${k}_{s0}$ and ${k}_{i0}$, and are related by the phase matching conditions (\ref{phasematch}) to each other 
and to the angles $\theta$.

It is of interest to point out that $F_x$ is a function of the angular variation differences, while $F_z$ is a function of their sum. Therefore (considering the opposite 
signs of $\theta_s$ and $\theta_i$ ) $F_x$ defines the range of angles over which both the signal and idler differ from the exact phase-matching directions, while 
preserving the angle between them. The function $F_z$ defines the range of variation of the angle between the signal and idler. This is similar to  factorization of the 
temporal (longitudinal) part of the biphoton amplitude, which has been shown to break up as a product of functions of the sum and difference of two detection times: 
$u(t_1-t_2)\times v(t_1+t_2)$ \cite{shih94c}. In this case, $v$ takes on the shape of the pump frequency spectrum (and is similar to $F_x$ in the present study) 
and $u$ depends essentially on the polarization and wavelength dispersion properties of the crystal  (and is similar to $F_z$ in the present study).

\begin{figure}[htp]
\centerline{
\input epsf
\setlength{\epsfxsize}{3in}
\setlength{\epsfysize}{4.1in}
\epsffile{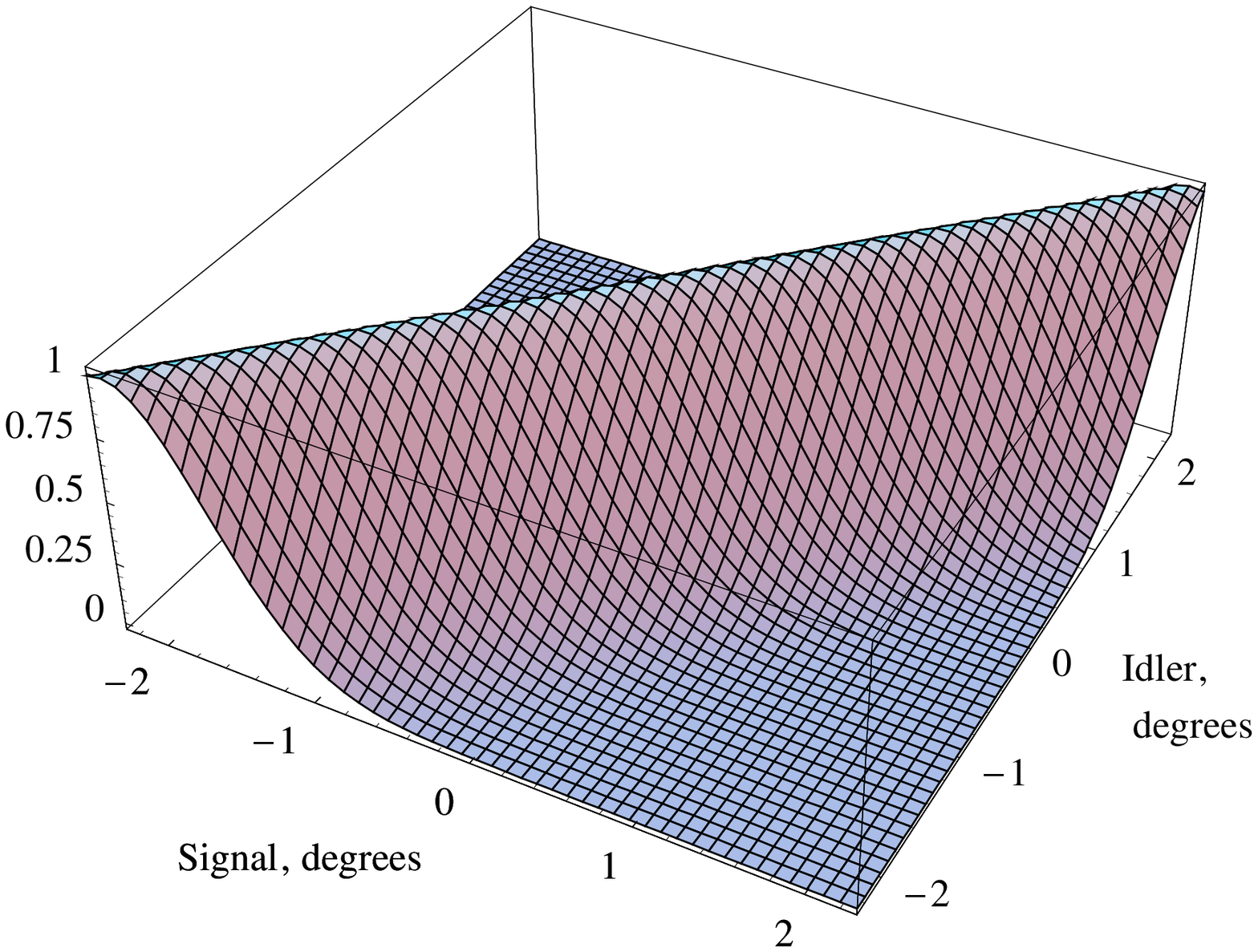}
\setlength{\epsfxsize}{3in}
\setlength{\epsfysize}{4.1in}
\epsffile{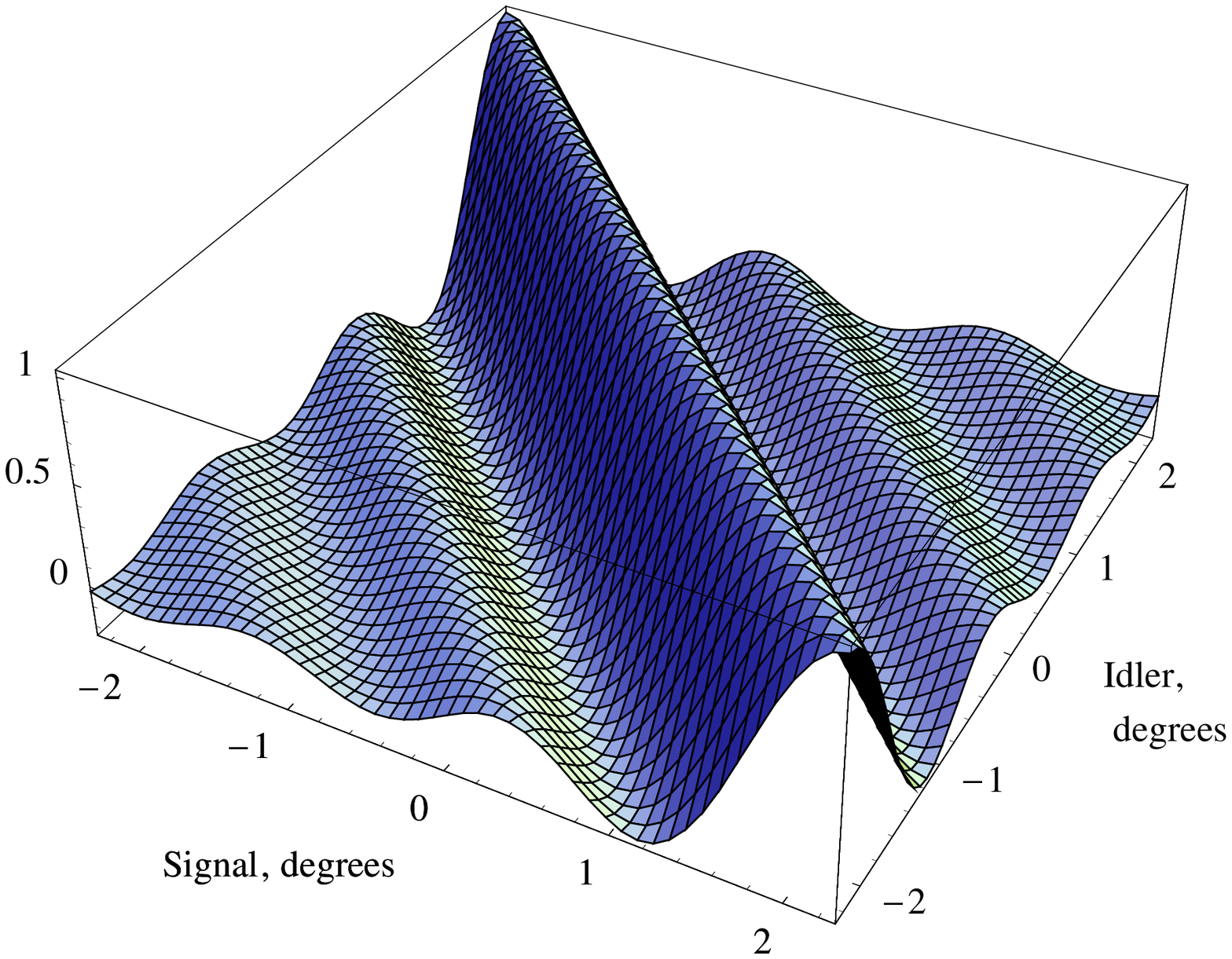}
}
\vspace*{-2in}
\caption[fs]{\label{fig:fs}The transverse part of the biphoton amplitude $F_x(0,\Delta\theta_s,\Delta\theta_i)$ (on the left) and 
$F_z(0,\Delta\theta_s,\Delta\theta_i)$ (on the right). The signal and idler wavelengths are  $\lambda_s = 690$ nm, $\lambda_i = 715$ nm.}
\end{figure}

\begin{figure}[htp]
\centerline{
\input epsf
\setlength{\epsfxsize}{2in}
\setlength{\epsfysize}{4in}
\epsffile{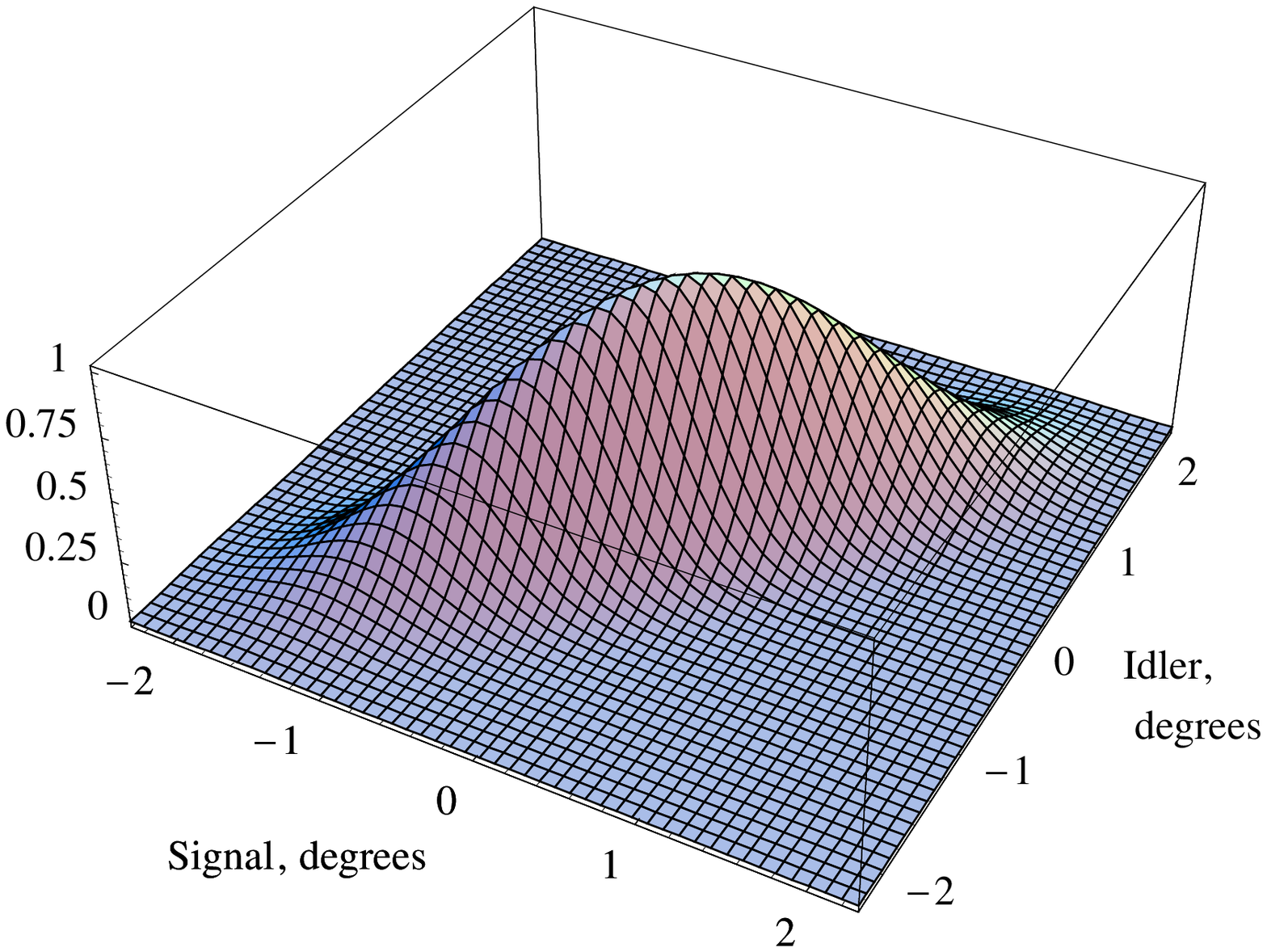}
\setlength{\epsfxsize}{2in}
\setlength{\epsfysize}{4in}
\epsffile{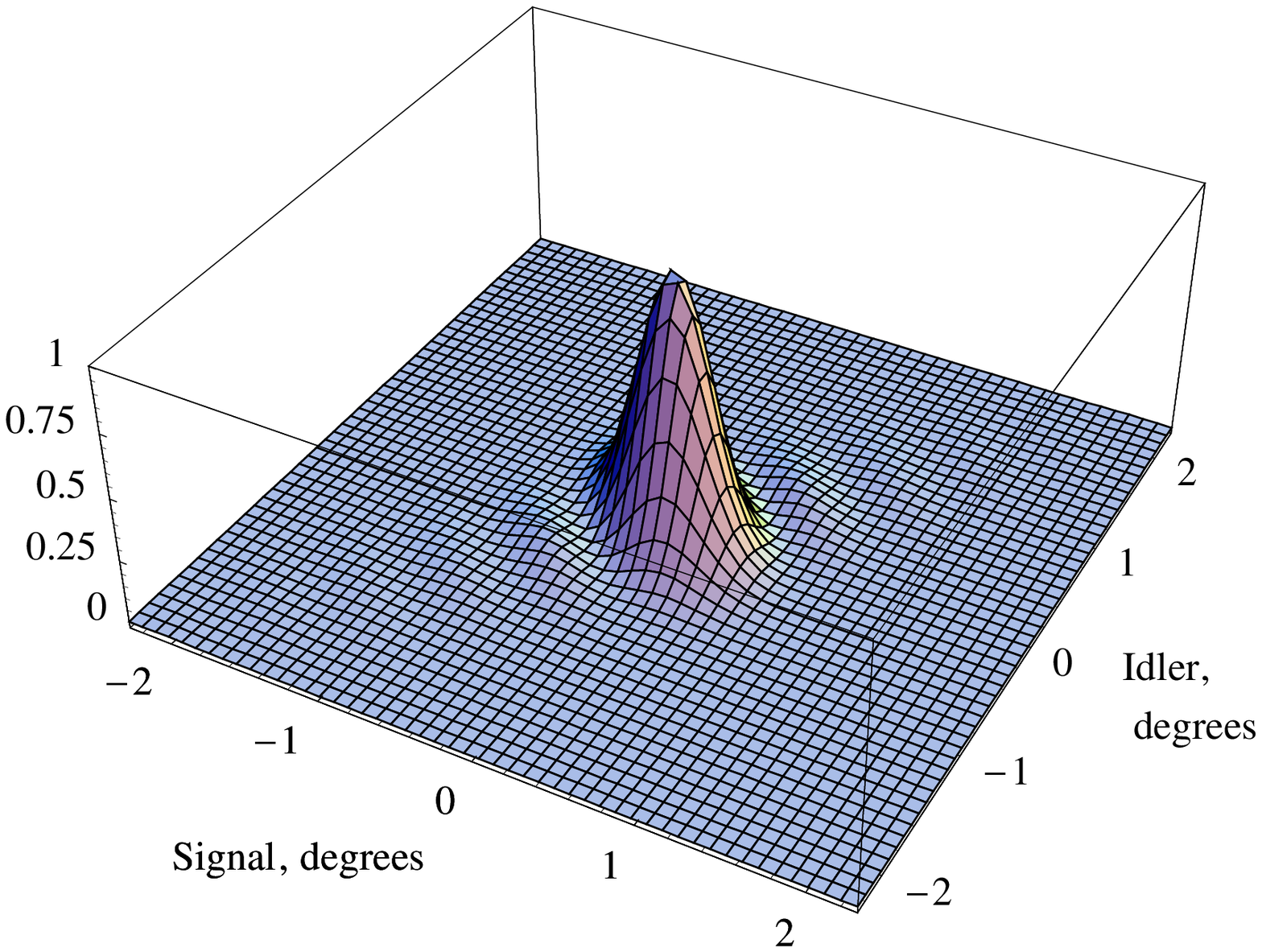}
\setlength{\epsfxsize}{2in}
\setlength{\epsfysize}{4in}
\epsffile{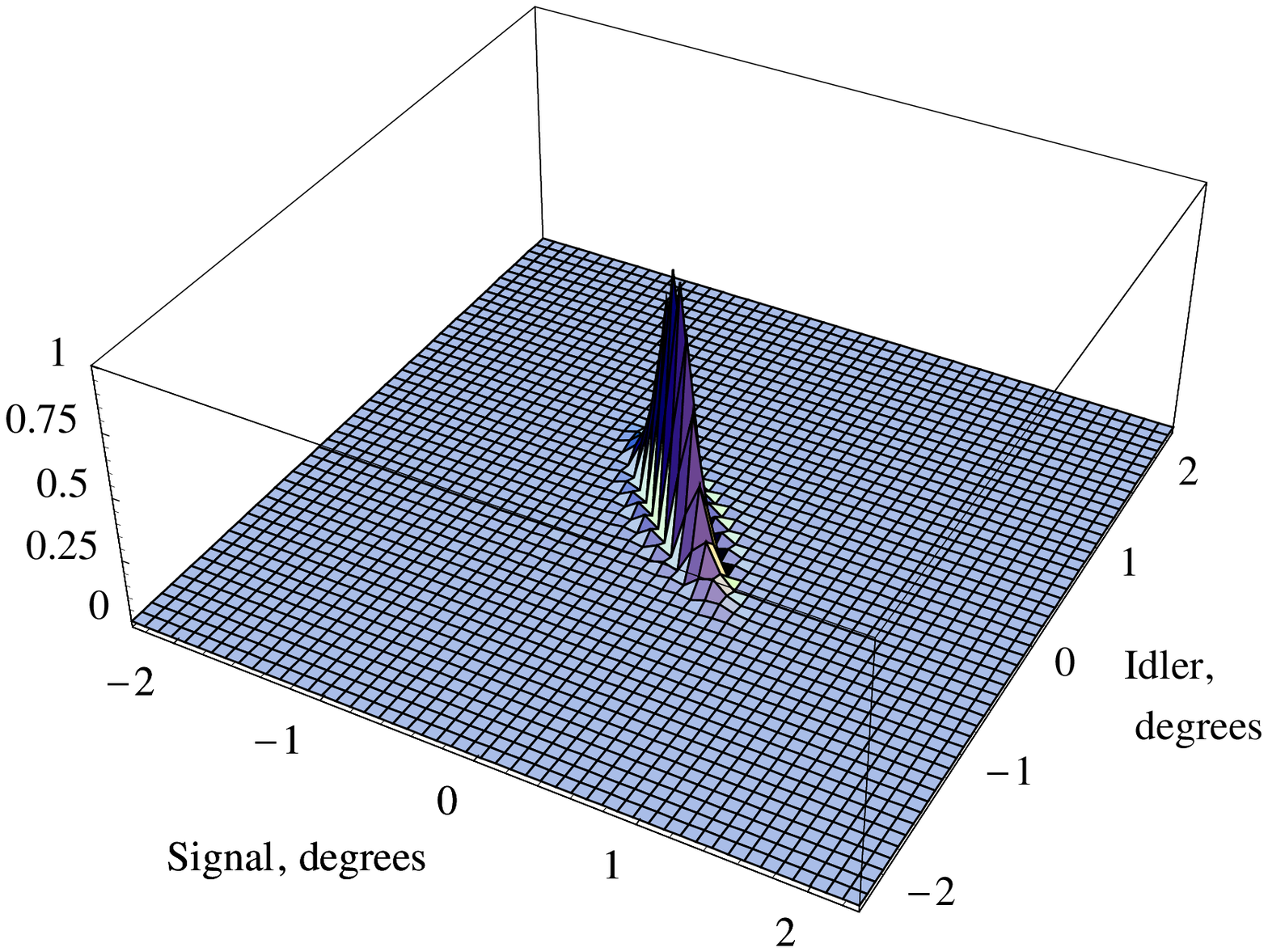}
}
\vspace*{-1.8in}
\caption{\label{fig:f}The biphoton transverse correlation functions $F^2(0,\Delta\theta_s,\Delta\theta_i)$. The signal and idler wavelengths are close to degeneracy 
(on the left); $\lambda_s = 690$ nm, $\lambda_i = 715$ nm (in the center); and $\lambda_s = 650$ nm, $\lambda_i = 754$ nm (on the right).}
\end{figure}

The internal angles in Eq. (\ref{fxfz}) may be replaced by the external ones via the Snell's law. The results are shown in Fig.\ref{fig:fs} for the pair $\lambda_s = 
690$ nm, $\lambda_i = 715$ nm. The product of $F_x(0,\Delta\theta_s,\Delta\theta_i)$ and $F_z(0,\Delta\theta_s,\Delta\theta_i)$ gives  
$F(0,\Delta\theta_s,\Delta\theta_i)$, whose absolute square can be interpreted as a correlation function of the angular variations $\Delta\theta_s$ and 
$\Delta\theta_i$. It turns out that the width of $F_z$ strongly depends on the signal and idler wavelength variation from degeneracy (when  $\lambda_s =\lambda_i 
$), getting broader as these wavelengths approach the degenerate values.  Three examples of $F^2(0,\Delta\theta_s,\Delta\theta_i)$ for different signal and idler 
wavelengths are shown in Fig.\ref{fig:f}, where the middle plot corresponds to the product of  functions from Fig.\ref{fig:fs}.

\begin{figure}[htp]
\centerline{
\input epsf
\setlength{\epsfxsize}{4in}
\setlength{\epsfysize}{4.6in}
\epsffile{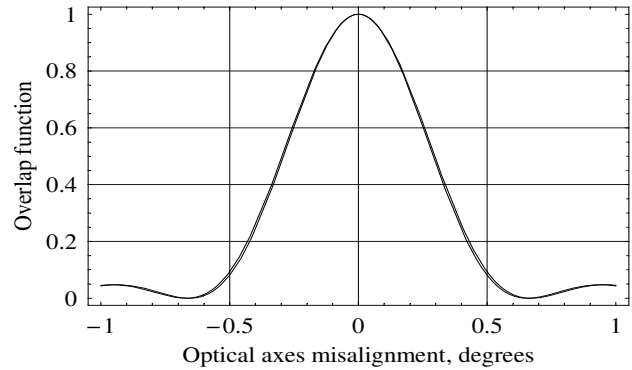}
}
\vspace*{-2.5in}
\caption{\label{fig:dalpha}The overlap function vs. misalignment of optical axes of two crystals.}
\end{figure}

Understanding of the transverse correlation properties of biphotons allows for a more thorough analysis of the biphoton up-conversion experiment. In particular, it 
allows us to establish the required degree of precision in alignment of the crystals' optical axes. To do this we substitute the arguments of the correlation function 
$F^2(0,\Delta\theta_s,\Delta\theta_i)$ with the phase-matching angles variations:
\begin{equation}
\Delta\theta_{s,i}= \frac{\partial\theta_{s,i}}{\partial\alpha}\Delta\alpha,
\label{dalpha}
\end{equation}
where the derivative is calculated with respect to the angle $\alpha$ between the optical axis and the pump beam. The function obtained from from substituting Eq. 
(\ref{dalpha}) into $F^2(0,\Delta\theta_s,\Delta\theta_i)$ represents the degree of overlap between the the down- and up-conversion phase matching conditions for 
crystals whose optical axes are misaligned by $\Delta\alpha$. In the following, we will call it the overlap function. Curiously, this function is practically uniform 
throughout the entire SPDC spectrum. Fig.\ref{fig:dalpha} shows two plots of this function: for degenerate, and for far non-degenerate wavelengths, that looks like a 
single line.

\begin{figure}[htp]
\centerline{
\input epsf
\setlength{\epsfxsize}{2.8in}
\setlength{\epsfysize}{4in}
\epsffile{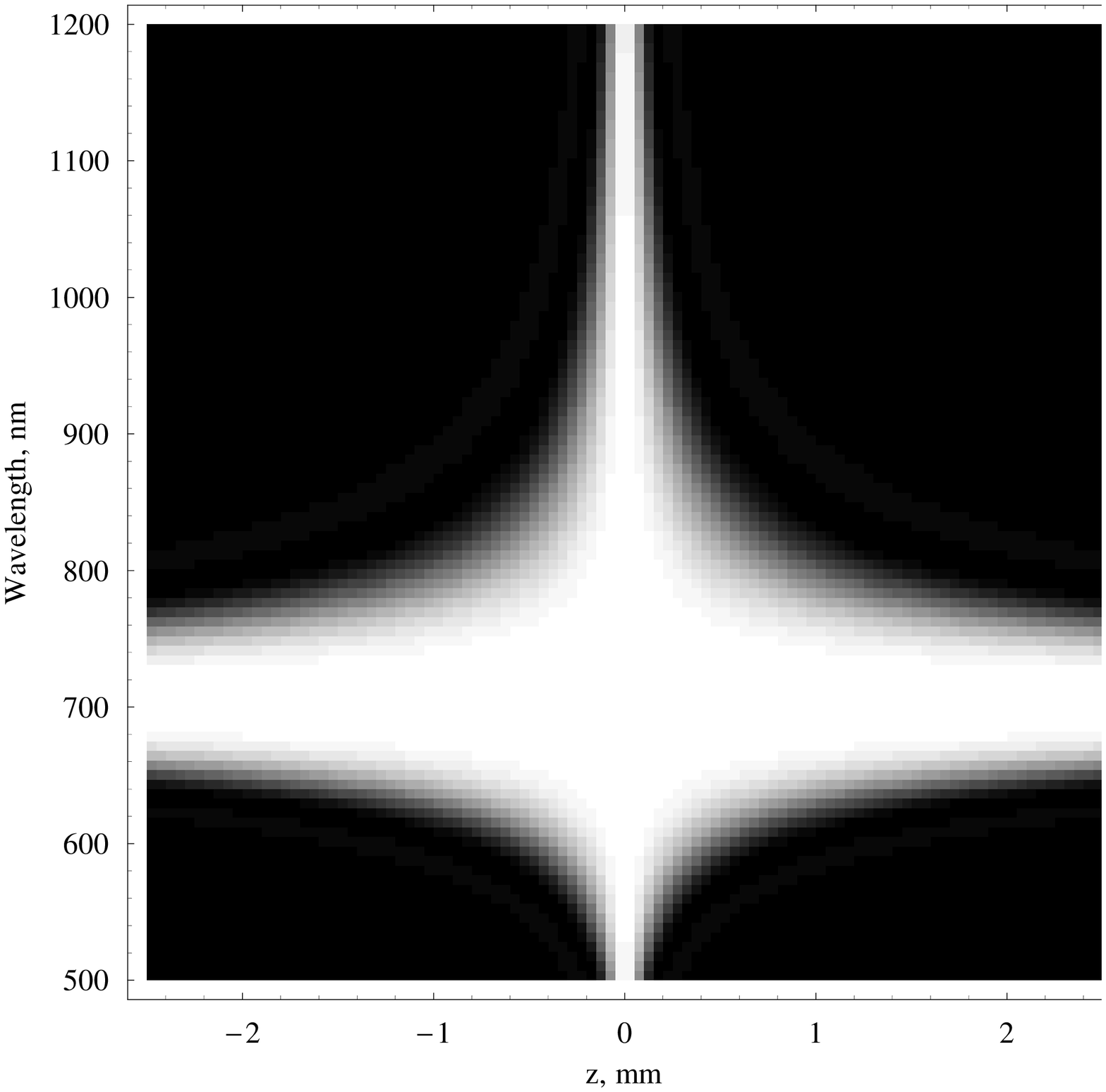}
\setlength{\epsfxsize}{3.2in}
\setlength{\epsfysize}{4.in}
\epsffile{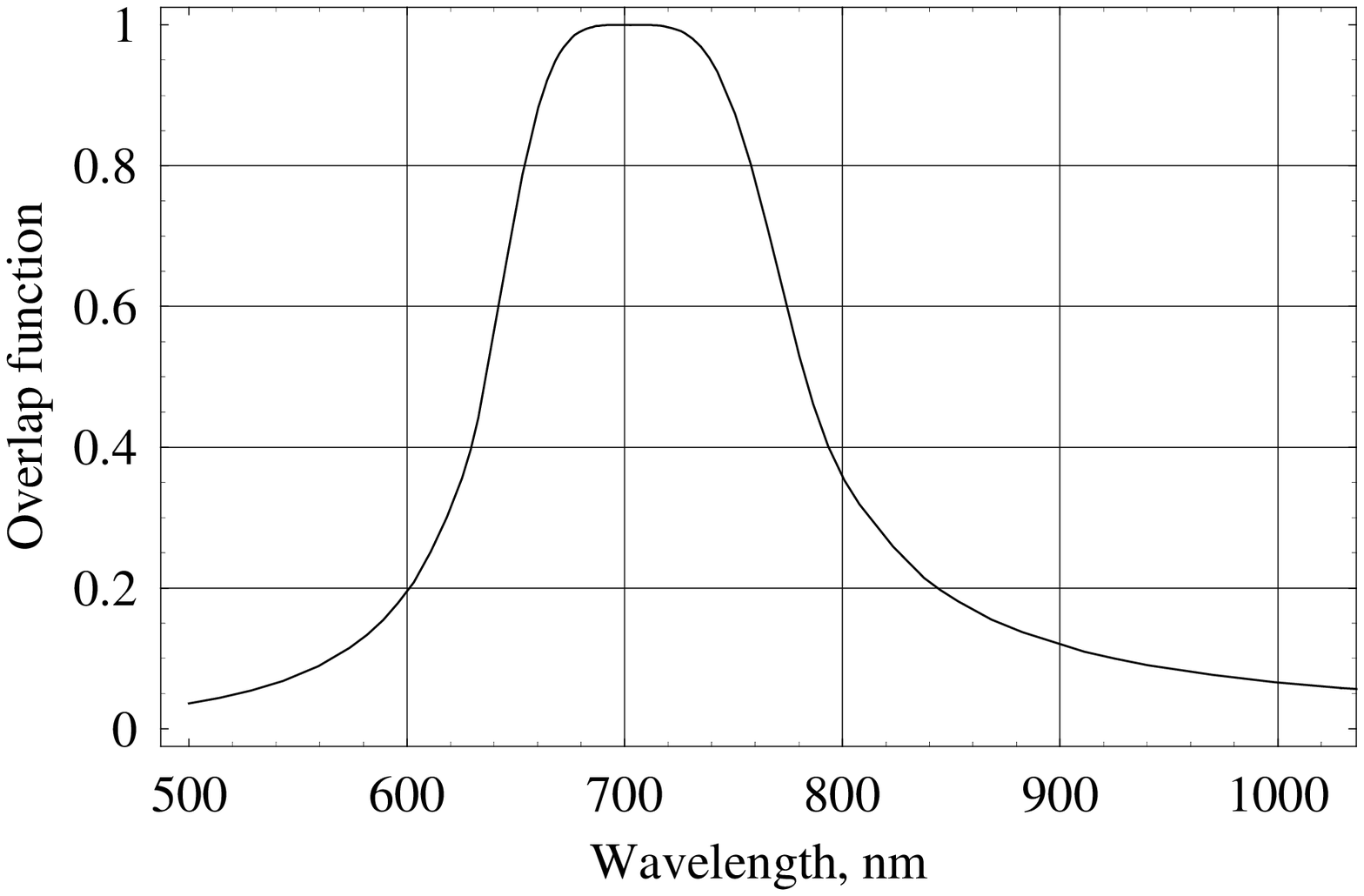}
}
\vspace*{-0.8in}
\caption{\label{fig:zefficiency}Left: the overlap function vs. the displacement $z$ from the imaging plane and the wavelength. Right: the overlap function for a 5 mm  
crystal vs. the wavelength.}
\end{figure}

From Fig.\ref{fig:dalpha} we see that the crystals' optical axes need to be aligned to about $0.2^\circ$ which is not hard to achieve. A bigger difficulty arises from 
the finite length of the crystals. Even though the parts of the down-converting crystal displaced from the 1:1 imaging plane still do get imaged onto the up-converting 
crystal, the signal and idler rays now cross at angles different from their phase matching angles $\theta_s,\theta_i$.  These angular errors can be found from simple 
geometrical considerations as functions of  the linear displacement $z$ from the 1:1 imaging plane. Substituting them into $F^2(0,\Delta\theta_s,\Delta\theta_i)$ we 
find the overlap function as a function of $z$ for different parts of the SPDC spectrum. Integrating this function over $z$, we find the overall spectral overlap 
function. From Fig.\ref{fig:zefficiency} we see that in case of a 5 mm BBO crystal at collinear degenerate phase matching, only a relatively narrow wavelength range 
can be efficiently up-converted. As a result, the above estimate of 40 photons per second is considerably reduced. These and other difficulties have so far precluded 
a convincing experimental demonstration of biphoton detection via optical up-conversion, however we plan to continue our work in this direction.

\section{Photo-electric processes in a biphoton field}

Another system interesting for the study of two-photon processes is a metal or semiconductor surface. If the red threshold of the external photo-electric effect in 
such a system corresponds to the total energy of the photon pair, one can expect a detectable photo-current corresponding to two-photon processes in the near 
absence of single-photon photo-current. The advantage of this system over the optical up-conversion is a much larger possible two-photon response cross-section; 
the disadvantage is that, in practice, such systems are very hard to characterize  due to their complexity and a large number of simultaneously occurring physical 
processes.

The two-photon response of the photocathodes in photo-multiplier
tubes (PMT) has been studied for use in the characterization of temporal properties of ultrafast laser pulses in which the peak
intensity is extremely large \cite{femtopulsesbk,hattori00femto}. Two-photon ionization may proceed by a direct two-photon channel via a virtual level, or by an 
indirect (cascade) process in which the electron may
reside in an intermediate state, such as the conduction
band or a deep trap, for some period of time. In many cases it may be beneficial to have an intermediate state in single-photon resonance with the light, because it 
increases the two-photon absorption rate. However, for the observation
of two-photon light from SPDC, long-lived intermediate states may lead to a signal from
uncorrelated photons, which is undesirable.  The purpose of this work is to characterize
a potential material for two-photon detection and to develop an understanding of the relevant physics for optimizing a fast two-photon detector. We are interested in 
observing the two-photon photo-electric effect from biphoton light, initiated by quantum-correlated photon pairs, rather then by absorption of statistically 
independent pairs of photons. For efficient discrimination between these two competing processes, we need to make sure the longest lifetime of the intermediate 
states is not much longer than the biphoton correlation time.

\begin{figure}[htp]
\centering
\includegraphics[width=5in]{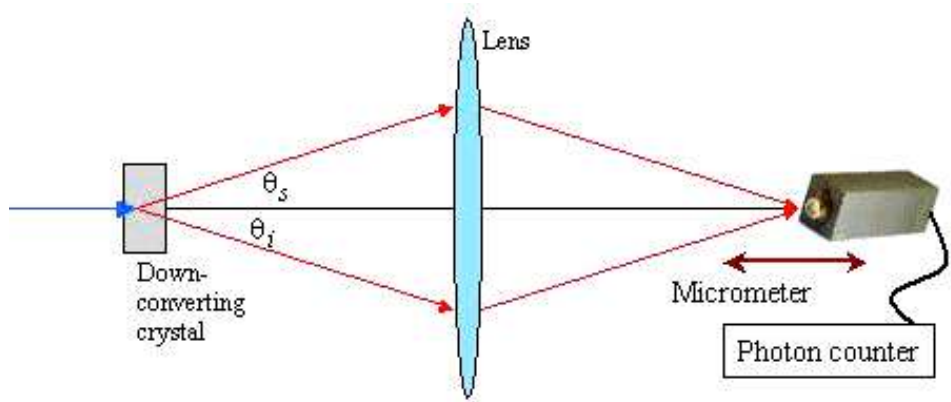}
\caption{Illustration of the two-photon photocathode experiment with biphoton light.} \label{cathodeimage}
\end{figure}

\begin{figure}[htp]
\centerline{
\input epsf
\setlength{\epsfxsize}{4.5in}
\setlength{\epsfysize}{2.8in}
\epsffile{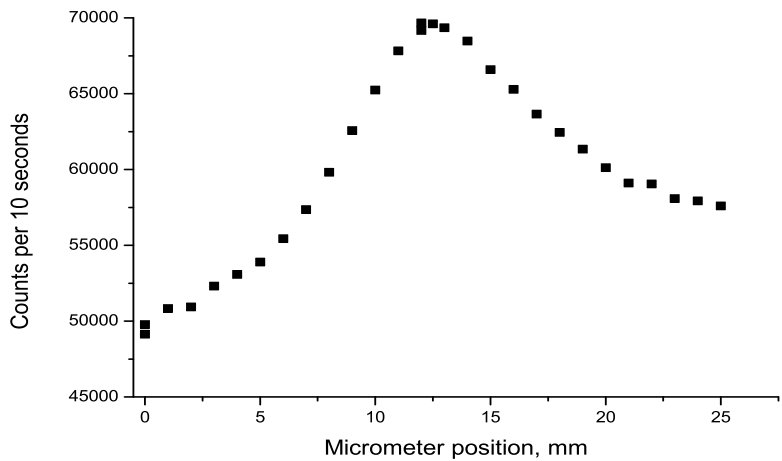}
}
\caption{\label{fig:two-phot_peak}Two-photon PMT response to the biphoton SPDC light as a function of the PMT position in Fig.\ref{cathodeimage}.}
\end{figure}

In our experiment, we image the biphoton source onto a ${\rm Cs_2Te}$ photocathode with diffraction-limited resolution, as illustrated in Fig.\ref{cathodeimage}. 
We use a PerkinElmer MH922P channel photomultiplier detector in photon counting mode. This CPM has a
${\rm Cs_2Te}$ cathode which has a bandgap of 3.3 eV and electron affinity
of less than 0.5 eV \cite{powell73cste}. The peak quantum efficiency, centered at 200 nm, is about 0.1. It falls off to approximately $10^{-3}$ at our pump 
wavelength (351.1 nm Argon Ion laser line), and to $10^{-8}$ at 702
nm, the degenerate signal and idler wavelength. Therefore one might expect to see a two-photon response from our  SPDC biphoton source. If the 
short-wavelength part of the SPDC spectrum is suppressed, the single-photon response is expected to be very small. In our experiment, this part of the spectrum, 
along with the UV pump, was suppressed by narrow-band and by  low-pass optical filters down to the dark noise of the PMT, which was at the level of a few 
counts per second. The PMT was moved in and out of the image plane. Although it always collected the same amount of light on its photocathode which, is 5 mm in 
diameter, all of the photon pairs are expected to focus to diffraction-limited spots only in the image plane. Therefore, the two-photon photo-electric signal should 
have a peak in the image plane. We have observed the expected behavior, see Fig.\ref{fig:two-phot_peak}. However, the origin of this peak turns out not to be due 
to the predicted two-photon effect, as discussed below.

Repeating the same experiment for attenuated CW laser light, we see a similar peak. Evidently, this peak arises from the fact that, in both the case of SPDC and the 
ordinary laser, the illuminated spot is the smallest, and therefore the intensity is the highest, in the image plane. Since the total light power is independent of the 
photocathode position, we obviously have some kind of a nonlinear (two-photon) process. This process, however, is very slow, which corresponds to a very large 
longitudinal extent of the two-photon detection volume. Physically, this most likely corresponds to a long lifetime of some intermediate state. As a result, a tight 
temporal (longitudinal) correlation of the biphoton field gives it no advantage over the attenuated classical light that has photons with Poissonian statistics.

\begin{figure}[htp]
\vspace*{0.8in}
\centerline{\hspace*{0.7in}
\input epsf
\setlength{\epsfxsize}{7in}
\setlength{\epsfysize}{5in}
\epsffile{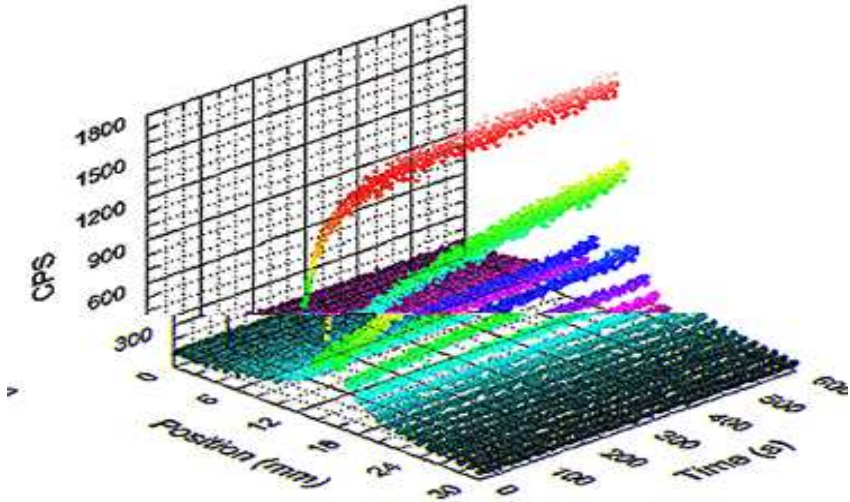}
}
\vspace*{-3in}
\caption{\label{fig:sensitization}Dependence of the two-photon counting rate (in counts per second) on the photocathode position with respect to the lens and on 
time.}
\end{figure}

Trying to determine the intermediate state lifetime, to characterize our prospective two-photon detector, we found that its two-photon sensitivity itself depends on 
time. This dependence is stronger for higher light intensities, that is, near the image plane. The situation can be described as the two-photon self-sensitization, which 
is illustrated in Fig.\ref{fig:sensitization} for an attenuated diode laser at around 650 nm. From Fig.\ref{fig:sensitization} we can see that the sensitization rate is higher 
for a higher intensity and that for the maximum light level we used (1.4 nW focused in a spot of 35 $\mu$m in diameter) the two-photon sensitivity can be increased 
by more than an order of magnitude over a period of time of about 100 seconds, at which point it saturates. We have also observed qualitatively similar 
power-dependent sensitization effect when keeping the illuminated area constant but changing the light power. However the preference was given to the other 
method because of experimental convenience.

\begin{figure}[htp]
\centering
\includegraphics[width=4in]{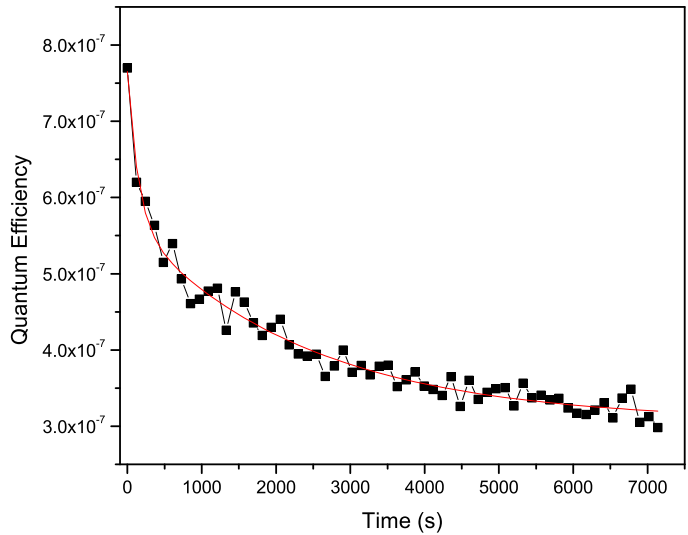}
\caption{Relaxation curve for a self-sensitized ${\rm Cs_2Te}$ photo cathode.} \label{relaxation}
\end{figure}

The photocathode remains saturated for some time after the light is turned off. Its relaxation back to the initial unsaturated sensitivity, as a function of time, is shown 
in Fig.\ref{relaxation}. To obtain this dependence, we have sensitized the photocathode at the maximum light level, and then let it sit in the dark, while periodically 
probing its response with weak light pulses. This response has been normalized to the probe light photon flux to yield directly the quantum efficiency. We found that 
the resulting decay curve fits to a bi-exponential function with decay constants of around 100 and 5 seconds, which may indicate the presence of at least two 
intermediate states. A possible explanation is that the electrons undergo indirect single-photon transition into deep trapped states in the bandgap. The nature of these 
states is unclear to us at this point. These could be surface states as well as localized impurity states, or the states associated with structural defects. In support of the 
deep-state hypothesis, we should mention the experimental observations of the photoconductivity dynamics in GaAs \cite{santic93photocond}. In this experiment, 
filling the deep traps by the  photo-activated electrons from a valence band results in a life-time increase of the electrons in the conduction band. The activation and 
relaxation time scales in this process are close to what we observe for ${\rm Cs_2Te}$.

\section{Conclusions and perspectives}

We have considered two types of two-photon processes that can be demonstrated with a faint biphoton field from SPDC and may in principle enable a fast 
two-photon detector: a process of coherent frequency up-conversion in an optically nonlinear media, and a two-photon photo-electric effect on a semiconductor 
photo-cathode. This is far from a complete list of processes that can be studied in this context. As the most obvious examples of the two-photon processes left 
beyond the scope of the present work, we would like to mention the internal photo-electric effect in broad-bandgap semiconductors; the earlier-mentioned 
ionization of alkali gases \cite{georgiades95,georgiades99}; and photo-chemical processes such as two-photon fluorescence and polymerization, e.g. 
\cite{cumpston99}.

Internal two-photon photo-electric processes should be similar to the external ones with the only difference that for the former the excited photo electrons remain in 
the conduction band and do not get emitted into vacuum. The internal processes can occur at much further distances from the surface than the external ones. This 
may simplify description of the internal processes by allowing one to neglect the effects of surface states and surface-bound impurities. We plan on studying this 
effect and expect that understanding of the mechanisms accompanying the external two-photon photo-electric effect will help.

Two-photon ionization of hydrogen-like gases is an example of a process that could be described analytically. Its simplicity is very attractive for research, however 
extremely low detection cross section, resulting from a low concentration of atoms in a gas, presents a serious experimental challenge in the case of a faint biphoton 
field. In photo-chemical fluorescence processes, on the other hand, the two-photon cross-section of an individual molecule can be quite large (reaching 
$1,250\times 10^{-50}$ cm$^4$ s per photon \cite{cumpston99}), while the concentration of molecules is also much larger than in gases. The two-photon 
polymerization is a subject of a great interest, since this is the key to the quantum two-photon lithography. While the two-photon polymerization of specially 
designed photoresists with strong laser pulses has been successfully demonstrated by several groups, e.g. \cite{witzgall98su8,cumpston99,kawata01bull}, the 
attempts to reproduce the same effect with biphoton field have not so far succeeded. While there are still several questions concerned with the photoresists 
properties (such as the reciprocity failure for long exposures, the role of thermal mechanisms, etc.) that need to be studied in order to achieve the success, the 
present study helps to answer one of the key questions by exploring the transverse correlation properties of the biphoton light. Specifically, for any particular imaging 
system, the transverse correlation function $F(0,\Delta\theta_s,\Delta\theta_i)$ that we have studied can be converted from the angular to linear variables 
$F(0,\Delta\rho_s,\Delta\rho_i)$ which has a direct relevance for calculating the two-photon exposure doses in quantum lithography.

{\bf Acknowledgments }

This work was carried out at the Jet Propulsion Laboratory, California Institute of Technology, under a contract with the National
Aeronautics and Space Administration.  We would also like to acknowledge financial support from the Office of Naval Research, the
National Security Agency, the National Reconnaissance Office, the Advanced Research and Development Activity, and the Defense
Advanced Research Projects Agency. We appreciate fruitful discussions with S.P. Kulik and A.N. Penin.

\bibliography{titles}

\begin{thebibliography}{10}

\bibitem{klyshko82}
D.N. Klyshko.
\newblock Transverse photon bunching and two-photon processes in the field of
  parametrically scattered light.
\newblock {\em Sov. Phys. JETP}, {\bf 56}:753--59, (1982).

\bibitem{Giord65}
J.A. Giordmaine and R.C. Miller.
\newblock Tunable coherent parametric oscillation in {L}i{N}b{O}$_3$ at optical
  frequencies.
\newblock {\em Phys. Rev. Lett.}, {\bf 14}:973--6, (1965).

\bibitem{burlakov97}
A.V. Burlakov, M.V. Chekhova, D.N. Klyshko, S.P. Kulik, A.N. Penin, Y.H. Shih,
  and D.V. Strekalov.
\newblock Interference effects in spontaneous two-photon parametric scattering
  from two macroscopic regions.
\newblock {\em Phys. Rev. A}, {\bf 56}:3214--25, (1997).

\bibitem{burlakov01}
A.V. Burlakov, M.V. Chekhova, O.A. Karabutova, and S.P. Kulik.
\newblock Biphoton interference with a multimode pump.
\newblock {\em Phys. Rev. A}, {\bf 63}:053801, (2001).

\bibitem{pittman96a}
T.B. Pittman, D.V. Strekalov, A.~Migdall, M.H.Rubin, A.V. Sergienko, and Y.H.
  Shih.
\newblock Can two-photon interference be considered interference of two
  photons?
\newblock {\em Phys. Rev. Lett.}, {\bf 77}:1917--20, (1996).

\bibitem{strekalov98}
D.V. Strekalov, T.B. Pittiman, and Y.H. Shih.
\newblock What we can learn about single photons in a two-photon interference
  experiment.
\newblock {\em Phys. Rev. A}, {\bf 57}(1):567--570, Jan (1998).

\bibitem{einstein35}
A.~Einstein, B.~Podolsky, and N.~Rosen.
\newblock Can quantum mechanical description of reality be considered complete?
\newblock {\em Phys. Rev.}, {\bf 35}:777--780, (1935).

\bibitem{pan00ghz}
Pan JW, Bouwmeester D, Daniell M, Weinfurter H, and Zeilinger A.
\newblock Experimental test of quantum nonlocality in three-photon
  greenberger-horne-zeilinger entanglement.
\newblock {\em Nature}, {\bf 403}:515--519, (2000).

\bibitem{Bouwmeester97teleport}
Bouwmeester D, Pan JW, Mattle K, Eibl M, Weinfurter H, and Zeilinger A.
\newblock Experimental quantum teleportation.
\newblock {\em Nature}, {\bf 390}:575--579, (1997).

\bibitem{kim01teleport}
Kim YH, Kulik SP, and Shih YH.
\newblock Quantum teleportation of a polarization state with a complete bell
  state measurement.
\newblock {\em Phys. Rev. Lett.}, {\bf 86}(7):1370--1373, Feb (2001).

\bibitem{ekert91a}
A.K. Ekert.
\newblock Quantum cryptography based on {B}ell's theorem.
\newblock {\em Phys. Rev. Lett}, {\bf 67}:661--663, (1991).

\bibitem{Yuen86}
Yuen HP.
\newblock Generation, detection, and application of high-intensity
  photon-number-eigenstate fields.
\newblock {\em Physical Review Letters}, {\bf 56}:2176--79, (1986).

\bibitem{Yurke86}
Yurke B, McCall SL, and Klauder JR.
\newblock {S}{U}(2) and {S}{U}(1,1) interferometers.
\newblock {\em Phys. Rev. A}, {\bf 33}:4033--54, (1986).

\bibitem{lugiato02qimaging}
L.A. Lugiato, A.~Gatti, and E.~Brambilla.
\newblock Quantum imaging.
\newblock {\em J. Opt. B}, {\bf 4}:S1--S8, (2002).

\bibitem{jozsa00clock}
R.~Jozsa, D.S. Abrams, J.~P. Dowling, and C.~P. Williams.
\newblock Quantum clock synchronization based on shared prior entanglement.
\newblock {\em Phys. Rev. Lett.}, {\bf 85}:2010--13, (2000).

\bibitem{Giovannetti02clock}
V.~Giovannetti, S.~Lloyd, and L.~Maccone.
\newblock Positioning and clock synchronization through entanglement.
\newblock {\em Phys. Rev. A}, {\bf 65}:022309, (2002).

\bibitem{boto00}
A.N. Boto, P.~Kok, D.S. Abrams, S.L. Braunstein, C.~P. Williams, and J.P.
  Dowling.
\newblock Quantum interferometric optical lithography: exploiting entanglement
  to beat the diffraction limit.
\newblock {\em Phys. Rev. Lett.}, {\bf 85}:2733--36, (2000).

\bibitem{kok01}
P.~Kok, A.N. Boto, D.S. Abrams, C.~P. Williams, S.~L. Braunstein, and J.~P.
  Dowling.
\newblock Quantum interferometric optical lithography: Towards arbitrary
  two-dimensional patterns.
\newblock {\em Phys. Rev. A}, {\bf 63}:063407, (2001).

\bibitem{Bjork01}
Bjork G, Sanchez-Soto LL, and Soderholm J.
\newblock Entangled-state lithography: Tailoring any pattern with a single
  state.
\newblock {\em Phys. Rev. Lett.}, {\bf 86}:4516--19, (2001).

\bibitem{dangelo01}
M.~D'Angelo, M.V. Chekhova, and Y.H. Shih.
\newblock Two-photon diffraction and quantum lithography.
\newblock {\em Phys. Rev. Lett.}, {\bf 87}:013602, (2001).

\bibitem{strekalov02litho}
D.V. Strekalov and J.P. Dowling.
\newblock Two-photon interferometry for high-resolution imaging.
\newblock {\em J. of Mod. Opt.}, {\bf 49}:519--527, (2002).

\bibitem{gilbert00}
G.~Gilbert and M.~Hamrick.
\newblock Practical quantum cryptography: A comprehensive analysis (part one).
\newblock {\em quant-ph/0009027}, (2000).

\bibitem{abram86}
I.~Abram, R.K. Raj, J.L. Oudar, and G.~Dolique.
\newblock Direct observation of the second-order coherence of parametrically
  generated light.
\newblock {\em Phys. Rev. Lett.}, {\bf 57}:2516--2519, (1986).

\bibitem{georgiades95}
N.~Ph. Georgiades, E.~S. Polzik, K.~Edamatsu, and H.~J. Kimble.
\newblock Nonclassical excitation for atoms in a squeezed vacuum.
\newblock {\em Phys. Rev. Let.}, {\bf 19}, (1995).

\bibitem{georgiades99}
N.P. Georgiades, E.S. Polzik, and H.J. Kimble.
\newblock Quantum interference in two-photon excitation with squeezed and
  coherent fields.
\newblock {\em Phys. Rev. A}, {\bf 59}:676--690, (1999).

\bibitem{perinabk}
J.~Perina.
\newblock {\em Coherence of light}.
\newblock D. Reidel Publishing Comp., 2nd edition, (1985).

\bibitem{klyshkobk}
D.N. Klyshko.
\newblock {\em Photons and Non-linear Optics}.
\newblock Gordon and Breach Science, New York, 1988.

\bibitem{shih88}
Y.H. Shih and C.O. Alley.
\newblock New type of {E}instein-{P}odolsky-{R}osen experiment using pairs of
  light quanta produced by optical parametric down conversion.
\newblock {\em Phys. Rev. Lett.}, {\bf 61 }:2921--2924, (1988).

\bibitem{hong87}
C.K. Hong, Z.Y. Ou, and L.~Mandel.
\newblock Measurement of subpicosecond time intervals between two photons by
  interference.
\newblock {\em Phys. Rev. Lett.}, {\bf 59}:2044--2046, (1987).

\bibitem{rubin96}
M.H. Rubin.
\newblock Transverse correlation in optical spontaneous parametric down
  conversion.
\newblock {\em Phys. Rev. A}, {\bf 54}:5349, (1996).

\bibitem{abouraddy01imaging}
A.F. Abouraddy, B.E.A. Saleh, A.V. Sergienko, and M.C. Teich.
\newblock Role of entanglement in two-photon imaging.
\newblock {\em Phys. Rev. Lett.}, {\bf 87}:123602, (2001).

\bibitem{nasr02biphoton}
M.B. Nasr, A.F. Abouraddy, M.C. Booth, B.E.A. Saleh, A.V. Sergienko, M.C.
  Teich, M.~Kempe, and Ralf Wolleschensky.
\newblock Biphoton focusing for two-photon excitation.
\newblock {\em Phys. Rev. A}, {\bf 65}:023816, (2002).

\bibitem{shih94c}
Y.H. Shih, A.V. Sergienko, M.H. Rubin, T.E. Kiess, and C.O. Alley.
\newblock Two-photon entanglement in type-{I}{I} parametric down-conversion.
\newblock {\em Phys. Rev. A}, {\bf 50}:23--28, (1994).

\bibitem{femtopulsesbk}
C.Rulliere, editor.
\newblock {\em Femtosecond Laser Pulses, Principles and Experiments}.
\newblock Springer-Verlag, (1998).

\bibitem{hattori00femto}
T.~Hattori, Y.~Kawashima, M.~Daikoku, H.~Inouye, and H.~Nakatsuka.
\newblock Femtosecond two-photon response dynamics of photomultiplier tubes.
\newblock {\em Jpn. J. Appl. Phys.}, {\bf 39}:4793--98, (2000).

\bibitem{powell73cste}
R.~A. Powell, W.~E. Spicer, G.~B. Fisher, and P.~Gregory.
\newblock Photoemission studies of cesium telluride.
\newblock {\em Phys. Rev. B}, {\bf 8}:15, (1973).

\bibitem{santic93photocond}
B.~Santic, U.V. Desnica, N.~Radic, D.~Desnica, and M.~Pavlovic.
\newblock Photoconductivity transients and photosensitization phenomena in
  semiinsulating {G}a{A}s.
\newblock {\em J. of Appl. Phys.}, {\bf 73}:5181--84, (1993).

\bibitem{cumpston99}
B.H.~Cumpston {\emph et al.}
\newblock Two-photon polymerization initiators for three-dimentional optical
  data storage and microfabrication.
\newblock {\em Nature}, {\bf 398}:51--54, (1999).

\bibitem{witzgall98su8}
G.Witzgall, R.~Vrijen, E.~Yablonovitch, V.~Doan, and B.J. Schwartz.
\newblock Single-shot two-photon exposure of commercial photoresist for
  production of three-dimentional structures.
\newblock {\em Optics Letters}, {\bf 23}:1745--47, (1998).

\bibitem{kawata01bull}
S.~Kawata, H.-B. Sun, T.~Tanaka, and K.~Takada.
\newblock Finer features for functional microdevices.
\newblock {\em Nature}, {\bf 412}:697--98, (2001).

\end{thebibliography}
\end{document}